# Synthesis of large single-crystal hexagonal boron nitride grains on Cu-Ni alloy

Guangyuan Lu[1,*], Tianru Wu[1,*], Qinghong Yuan[2], Huishan Wang[1,3], Haomin Wang[1], Feng Ding[4], Xiaoming Xie[1,5], Mianheng Jiang[1,5]

[1]State Key Laboratory of Functional Materials for Informatics, Shanghai Institute of Microsystem and Information Technology, Chinese Academy of Sciences, 865 Changning Road, Shanghai 200050, China

[2] Department of Physics, East China Normal University, Shanghai 200241, China

[3]School of Physics and Electronics, Central South University, Changsha 410083, China

[4]Institute of Textiles and Clothing, Hong Kong Polytechnic University, Kowloon, Hong Kong 999077, China

[5]School of Physical Science and Technology, Shanghai Tech University, 319 Yueyang Road, Shanghai 200031, China

[*]These authors contributed equally to this work.

Correspondence and requests for materials should be addressed to X. X.
 (email: xmxie@mail.sim.ac.cn)

**Abstract:**

Hexagonal boron nitride (h-BN) has attracted significant attention due to its superior properties as well as its potential as an ideal dielectric layer for graphene-based devices. The h-BN films obtained via chemical vapor deposition in earlier reports are always polycrystalline with small grains due to high nucleation density on substrates. Here we report the successful synthesis of large single-crystal h-BN grains on rational designed Cu-Ni alloy foils. It is found that the nucleation density can be greatly reduced to 60 per $mm^2$ by optimizing Ni ratio in substrates. The strategy enables the growth of single-crystal h-BN grains up to 7,500 $\mu m^2$, ~ 2 orders larger than that in previous reports. This work not only provides valuable information for understanding h-BN nucleation and growth mechanisms, but also gives an effective alternative to exfoliated h-BN as a high-quality dielectric layer for large-scale nano-electronic applications.

Hexagonal boron nitride (h-BN) is a representative two-dimensional (2D) crystal with a band gap of about 6 eV (ref. 1). Due to the strong covalent sp$^2$ bonds in the plane, it has very high mechanical strength, thermal conductivity, and chemical stability[2-6]. The h-BN layer with atomically smooth surface free of dangling-bonds/charge-traps makes it an ideal dielectric substrate for field effect transistor[7] and spacer for novel tunneling devices[8,9]. The characters of h-BN make it possible to realize ballistic electronics at room temperature[10,11]. Normally, uniform crystalline h-BN flakes are mostly obtained by mechanical exfoliation from h-BN single crystals[7,11-15]. But the flakes are obtained through a highly skilled manual process--mechanically exfoliation and transferring. It is not a scalable method for practical applications. So far, many efforts have been taken on various substrates such as Ni[16-21], Cu[3,21-27], Pt[28,29], Ru[30,31] and Co[32] to obtain large h-BN crystal via chemical vapor deposition (CVD) process. However, those grains in h-BN films are very small (usually <50 μm$^2$) because of high nucleation density at the early growth stages[10,11,15-17,19,22]. The small grains lead to high density of grain boundaries and dangling bonds, which are known as structural defects in h-BN[33,34]. The defects definitely give rise to higher surface roughness / charge impurities, and then degrade the performance of electronic devices made on it[35]. Therefore, the growth of high-quality large h-BN crystal is essential in the commercial-level manufacture for its further application in electronic, photonic and mechanical devices.

Here we report the successful synthesis of large–size single-crystal h-BN grains on rational designed binary Cu-Ni alloy. It is observed that under our growth conditions the nucleation density can be dramatically decreased by introducing Ni to the Cu substrate. 10-20 atom % of Ni can reduce the nucleation density down to 60 per mm$^2$, and h-BN grains of ~7,500 μm$^2$ in area can be obtained. The grains are about two orders larger than those in previous reports[10,11,15-17,19,22]. The introduction of Ni is also proved to enhance the decomposition of poly-aminoborane and help synthesize high-quality h-BN grains free of nanoparticles. The characterizations such as AES, AFM and TEM confirm that h-BN grains are uniform. Our results demonstrate that the CVD-grown h-BN layer with large grain can help effectively preserve the intrinsic properties of graphene, and will certainly facilitate the further application of wafer-scale electronics.

## Results

**h-BN growth.** In our work, Cu-Ni alloy was used as the catalytic substrate for h-BN growth. The alloy was obtained by annealing Ni-coated Cu foils at 1,050 °C for 2 hours in flowing hydrogen. Ammonia borane ($H_3BNH_3$, also called borazane) was supplied as the precursor for h-BN growth. By heating the precursor to 70-90 °C, borazane began to dissociate and the products were carried into the reaction chamber in a $H_2$ flow at a pressure of about 50 Pa. The growth temperature was about 1,050 °C. The schematic of process flow is illustrated in Fig. 1a, and the experimental details are given in Supplementary Fig. 1.

Fig. 1b-1e show the scanning electron microscopy (SEM) images of h-BN grown at 1,085 °C on Cu-Ni alloy substrates with 15 atom % Ni. It can be clearly seen that the h-BN grains nucleate, grow up and finally coalesce into a continuous film. The typical lengths of side for triangle h-BN grains after 10 min and 40 min of growth were 5 μm and 60 μm, respectively, while grains with the length of side up to 130 μm were obtained after 60 min. All the h-BN grains obtained are in triangular shape (Fig. 1b-1d), which implies that these grains are all single crystals with nitrogen-terminated zigzag edges[16,17,23]. h-BN grains growing across alloy grain boundaries are also frequently observed, suggesting that under the conditions h-BN growth is surface-mediated[23].

Experiments were carried out to understand the influence of Ni concentration on h-BN growth. Fig. 2a shows a typical SEM image of h-BN grains grown on pure Cu foil for 10 min at 1,050 °C. It could be observed that the nucleation density of the h-BN grains is so high that the triangular-shaped grains with only ~10-15 μm in the length of side are about to coalesce. Fig. 2b-2d show SEM images of h-BN grains grown for 60 min on Cu-Ni alloy foils with 10, 20, and 30 atom % Ni, respectively. And the relationship between the largest grain size and growth time is given in Fig. 2e, while the relationship between the number of grains and growth time is shown in Fig. 2f. It can be found that the nucleation density decreased notably after introducing Ni element. On Cu-Ni alloy substrate with 10-20 atom % Ni, the largest length of side for single-crystal h-BN grains reaches about 70-90 μm with a lateral growth rate of ~1.2 μm min$^{-1}$. The nucleation density dropped further on 30 atom % Ni substrates, but the lateral growth rate dropped to ~0.3 μm min$^{-1}$, with the largest length of side for grains only achieved about 30 μm. The same experiments were also taken on Ni foils and Cu-Ni alloy foils with 50 atom % Ni, but in the present process window, h-BN grains were not observed even after 90 min of growth for both cases (see Supplementary Fig. 2).

The growth mechanism of h-BN on Cu is mainly surface-mediated[23,25], while on Ni, it is neither surface-limited nor dominated by segregation and precipitation of B and N, but rather depends on surface chemistry of Ni-B and Ni-N[18]. Our experimental parameters of h-BN growth on Cu-Ni substrate with 10-20 atom % Ni are based on the surface-mediated mechanism on Cu foils. With further increase of Ni content, new growth mechanism based on solid-gas reactions involving Ni-B and Ni-N sets in, so that the growth of h-BN would move entirely to a different process window, accounting for decreased growth rate or even inability for phase formation.

As shown in Fig. 1b-1d and Fig. 2f, it is also discovered that nucleation continued with its rate increased nonlinearly with time while the early nuclei developed in size during growth. Similar phenomenon has also been reported for h-BN growth on Cu foils[23]. It is therefore clear that the key to success for large grain size is to reduce the initial nucleation density and to suppress the continuous nucleation during growth.

**Mechanisms.** Figure 3a-3d compare the typical surface morphologies of h-BN grown on Cu (10 min at 1,050 °C) and Cu-Ni alloys (25 min at 1,050 °C, 5-20 atom% Ni). Many nanoparticles (appearing white in SEM) are observed on Cu surface, while the density decreases dramatically with the increasing of Ni concentration in substrates. The nanoparticles disperse randomly amongst h-BN grains or accumulate along its perimeters, and are statically correlated to the h-BN grain nucleation (see more information in Supplementary Fig. 3).

As illustrated in Figure 3e, the borazane precursor ($H_3BNH_3$) decomposes into gaseous borazine ($B_3N_3H_6$) and solid poly-aminoborane ($BH_2NH_2$)$_n$ upon heating to 70~90 °C (refs. 23,24,36-40). Poly-aminoborane can hardly develop into h-BN unless a very high temperature of 1,170-1,500 °C is applied if without catalytic substrates[40]. Previous studies indicate that on catalytic surface the resulting polymer undergoes a two-stage weight loss process: two-dimensional cross-linking reaction of B-H and N-H groups is initiated from 125 to 200 °C, and then dehydrogenation from unaligned chain branches continues from 600 °C to 1,100 °C (refs. 20,36). The reactions involved suggest that the nanoparticles are most likely a complex mixture of poly-aminoborane and its partially dehydrogenated derivatives, consistent with earlier reports[23,24,26,41] and our Auger investigations (Supplementary Fig. 3a-e).

To have a better understanding on the suppressed formation of nanoparticles at higher Ni content, we carried out simplified density functional theory (DFT) calculations of decomposition energy of $(BH_2NH_2)_{n=1}$ on substrates with different compositions (Fig. 3f, details of the calculations are given in Supplementary Fig. 4). The calculation shows that dehydrogenation of $BH_2NH_2$ becomes easier by increasing the Ni content in our alloy substrates in a certain range. On the other hand, the decomposition energy of BN, dehydrogenated derivative of $BH_2NH_2$, drops more evidently by introducing Ni to the Cu substrate. It is apparent that the introduction of Ni can enhance the decomposition of poly-aminoborane, which helps the reactions of desorption or the formation of Ni-B and Ni-N phases[18], consistent with the weak h-BN growth on Cu-Ni alloy with 30 atom % Ni. To enable growth of h-BN on Cu-Ni alloy with high Ni content, one would have to increase the precursor supply, thus the growth of h-BN enters into a completely different process window[18,19,21,23,24].

**Characterization.** Fig. 4a-4c show the SEM image and the corresponding B(KLL) and N(KLL) Auger electron maps of h-BN grains grown on Cu-Ni alloy foil with 15 atom % Ni. It is found that distributions of B and N elements are homogeneous in the grains. Fig. 4d presents the AES spectra taken in the blue and magenta dotted areas shown in Fig. 4a, while the insets show the corresponding B(KLL) and N(KLL) spectra, respectively. The B(KLL) and N(KLL) peaks measured on h-BN grain are at 176 and 385 eV, respectively, in agreement with those given in previous reports[42,43]. It can also be seen that the BN-coated area (blue dot) shows a much weaker peak of O(KLL) than the uncoated area (magenta dot), proving the high oxidation resistance of the h-BN film[44]. Besides homogeneous h-BN grains, some triangular-shaped h-BN ad-layers with size of more than 10 μm were occasionally observed (Supplementary Fig. 5). The edges of the ad-layers are parallel to those of the first layer, implying strict stacking orders of the h-BN multilayers.

Fig. 4e shows the atomic force microscopy (AFM) image of a corner of a triangular-shaped h-BN grain transferred onto $SiO_2$/Si substrate. The height of the transferred film is about 0.455 nm, consistent with that reported for single-layer h-BN [23,24,28,29]. Fig. 4f shows a typical transmission electron microscopy (TEM) image obtained at the edge of the h-BN film on a TEM grid, confirming that the h-BN prepared is monolayered. The selected area electron diffraction (SAED) pattern presented in the inset of Fig. 4f shows a set of characteristic 6-fold symmetric spots, indicating that the h-BN grain is well-crystallized. TEM images containing 2-4 atomic layers were also occasionally detected, as shown in Supplementary Fig. 6. More characterizations of the h-BN grains, including X-ray photoemission spectroscopy (XPS), optical image and optical absorption spectrum are presented in Supplementary 7.

**Application.** We fabricated graphene/h-BN hetero-structure to evaluate the quality of CVD-grown h-BN (the transferring method is given in Supplementary Fig. 8). Fig. 5a shows the optical image of graphene/h-BN hetero-structure on a 90 nm $SiO_2$/Si substrate, with the Raman spectra taken in different areas presented in Fig. 5b. The spectrum taken in the surface of h-BN (magenta dot) exhibits a characteristic peak at 1,370 cm$^{-1}$ due to the $E_{2g}$ phonon mode of h-BN[45], consistent with previous report on the Raman signature of mechanical exfoliated monolayer h-BN films[13]. The peak near 1,450 cm$^{-1}$ is from the third order transverse optic (TO) mode of the silicon substrate[46,47]. Both the spectra taken in the surfaces of graphene/SiO$_2$ (olive dot) and graphene/h-BN (violet dot) indicate that the graphene grain is well-crystallized monolayer with few defects[48,49].

Graphene/h-BN and contrastive graphene/SiO$_2$/Si Hall devices were then made to further characterize the electronic properties. Fig. 5c displays the resistivity curves of typical graphene/h-BN and graphene/SiO$_2$/Si back-gate field-effect transistors (FETs), with the calculated mobility read 6,213 and 3,625 cm$^2$V$^{-1}$s$^{-1}$, respectively. The effectiveness of the CVD-grown h-BN as dielectric substrate for graphene devices is thus well proved. Moreover, as given in Fig. 5d, through the graphene/h-BN Hall device the half-integer quantum Hall effect is also observed, further confirming the high quality of both h-BN and graphene grains. As continuous h-BN films containing more than one atomic layers were synthesized and used as substrates, the mobility of our CVD-grown graphene is further improved to ~10,000 cm$^2$V$^{-1}$s$^{-1}$ (see Supplementary Fig. 9).

## Discussion

In summary, we successfully grow uniform large single-crystal h-BN grains via CVD method. By rationally designing Cu-Ni alloy substrate, the nucleation density of h-BN can be greatly decreased to 60 per mm$^2$. Single-crystal h-BN grains with 7,500 μm$^2$ are obtained, which are two orders larger in area than previous reports. Characterizations such as AES, AFM and TEM confirm the uniformity of h-BN grains. Electrical transport measurements show that the CVD-grown h-BN layers exhibit excellent dielectric performance. The success brings in hope in the successive CVD growth of high quality h-BN, as well as fabrication of graphene/h-BN hetero-structure or super-lattices, which are important for both fundamental research as well as device applications.

## Methods

**Deposition of Ni film on a Cu foil.** A 8 cm × 8 cm Cu foil (25μm, 99.8%, Alfa -Aesar) was first electrochemically polished to reduce the surface roughness. The polishing solution consisted of 500 mL of water, 250 mL of ethanol, 250 mL of orthophosphoric acid, 50 mL of isopropyl alcohol, and 5 g of urea. After 90 seconds of electrochemical polishing with the current set at 10 A, the thickness of the Cu foil reduced to about 20 μm. Then the polished Cu foil was loaded into an annealing tube and annealed at 1,050 °C for 2 h in a mixed Ar/H$_2$ (400/100 sccm) flow under atmospheric pressure (AP) to further increase its surface flatness and grain size.

After annealing, the Cu foil was electroplated with a layer of Ni. The electrolytic solution was a mixture of 1 L of water, 280 g of NiSO$_4$·6H$_2$O, 8 g of NiCl$_2$·6H$_2$O, 4 g of NaF and 30 g of H$_3$BO$_3$. With the current density set at 0.01 A cm$^{-2}$, a Ni film with a certain thickness was deposited on the Cu foil at a rate of 200 nm min$^{-1}$.

**Growth of h-BN grains.** The Ni-coated Cu foils were loaded into a 5 cm diameter CVD fused quartz tube as shown in Supplementary Fig. 1a. The distance between precursor and substrates is about 60 cm. The annealing and growth process is shown in Supplementary Fig. 1b. The substrates were first heated to 1,050 °C and maintained for 2 h in a H$_2$ flow at a total pressure of about 5 kPa. After that, the two metals were completely inter-diffused to Cu-Ni alloy foils with the atomic proportions determined by the thickness of the deposited Ni film before annealing. For h-BN growth, the temperature was changed to the specified value in the range of 1,050-1,090 °C, and the total pressure was maintained at 50 Pa. At the same time, borazane precursor placed in an $Al_2O_3$ boat was heated to 70-90 °C by a heating lamp. After growth, both the heating furnace and the heating lamp were quickly cooled down to room temperature.

**Characterizations.** h-BN grains were characterized by scanning electron microscopy (SEM) (Zeiss Supra55, operated at 3 kV), Auger electron spectroscopy (AES) (PHI670, operated at 10 kV), atomic force microscopy (AFM) (Bruker Dimension Icon, contact mode), transmission electron microscopy (TEM) (FEI Tecnai G2, 200 kV), Raman spectroscopy (Renishaw, 514 nm laser wavelength), X-ray photoemission spectroscopy (XPS) (AXIS UltraDLD) and UV-Vis spectrophotometer (EV300).

**DFT calculation.** All the calculations are performed within the framework of density functional theory (DFT) as implemented in the Vienna Ab initio Simulation Package (VASP). Electronic exchange and correlation are included through the generalized gradient approximation (GGA) in the Perdew–Burke–Ernzerhof (PBE) form. The interaction between valence electrons and ion cores is described by the projected augmented wave (PAW) method and the energy cut off for the plane wave functions is 400 eV. All structures were optimized until the maximum force component on each atom is less than 0.02 eV Å$^{-1}$. The vacuum layer inside the super-cell is kept as large as 14 Å to avoid the interaction of the adjacent unit cell. The slab model is composed of 4×4 repeating unit cells and the 2×2×1 k-point mesh is used for the calculation.

The dehydrogenation energies, $E_{dh}$, of each dehydrogenated borazane on the surfaces are calculated by $E_{dh} = E(\text{Sub.-BN})+2\times E(H_2)–E(\text{Sub.})–E(H_2BNH_2)$, in which $E(\text{Sub.-BN})$ is the total energy of fully dehydrogenated borazane adsorbed on the substrate; $E(\text{Sub.})$ and $E(H_2BNH_2)$ are the energies of substrate and separated $H_2BNH_2$, respectively; and $E(H_2)$ is the energy of hydrogen molecule in the vacuum. The decomposition energies are calculated by $E_{dc} = E(\text{Sub.-B--N}) –E(\text{Sub.-BN})$, in which $E(\text{Sub.-B--N})$ and $E(\text{Sub.-BN})$ are the energies of separated B, N atoms adsorbed on the substrate and BN cluster adsorbed on the substrate, respectively.

**Fabrication and measurement of graphene Hall devices.** CVD-grown monolayer graphene grains with a typical size of 50 μm were transferred onto as-grown h-BN grains on Cu-Ni alloy, followed by the etching of Cu-Ni and then the graphene/h-BN grains were transferred onto a highly doped p-type silicon wafer with 300 nm $SiO_2$ capping layer. For comparison, some graphene grains were transferred directly onto $SiO_2$/Si substrate. After that, a standard electron beam lithography technique was employed to pattern the Hall bar structure. The contact electrodes (10 nm Au/50 nm Ti) were deposited through electron beam evaporation. Both electrical transport and magneto-transport measurements were carried out in a PPMS-9T system from Quantum Design.

## Acknowledgements

We thank X.F. Zhang for providing CVD-grown graphene on Cu, J. Chen, X.Y. Liu and Q.J. Sun for the help in device fabrications. The work in Shanghai Institute of Microsystem and Information Technology, Chinese Academy of Sciences is partially supported by the National Science and Technology Major Projects of China (Grant No. 2011ZX02707), Chinese Academy of Sciences (Grant No. KGZD-EW-303 and XDB04010500), CAS International Collaboration and Innovation Program on High Mobility Materials Engineering, the National Natural Science Foundation of China (Grant No. 11304337, 11104303, 11274333, 11204339 and 61136005), and the projects from Science and Technology Commission of Shanghai Municipality (Grant No. 12JC1410100 and 12JC1403900).

## Author contributions

M.J. and X.X. supervised the research work. T.W. conceived the project. T.W. and G.L. designed the experiments. G.L. performed h-BN growth, transfer and characterization. F.D. and Q.H. performed the DFT calculations. H.M.W. and H.S.W. fabricated the graphene electronic devices and carried out the transport measurements. T.W., X.X. and G.L. analyzed the experimental data and designed the figures. H.M.W., G.L., T.W., X.X. and M.J. co-wrote the manuscript and all authors contributed to the critical discussions of the manuscript.

**Competing financial interests:** The authors declare no competing financial interests.

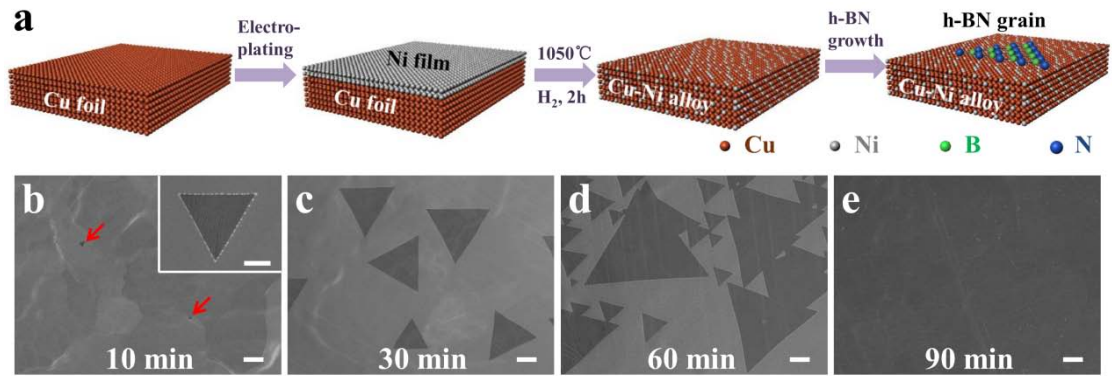

**Figure 1 | Rational design of Cu-Ni alloy substrates for h-BN growth.** (**a**) Schematic illustration showing the procedure of h-BN growth. (**b-e**) Typical SEM images of h-BN grains grown on Cu-Ni alloy foils with 15 atom % Ni at 1,085 °C for 10 min, 30 min, 60 min and 90 min, respectively. The red arrows in (**b**) show the sites of the h-BN grains, while the inset shows the enlarged image of one as-grown h-BN grain. The scale bars in (**b-e**) are 20 μm, and in inset in (**b**) is 2 μm.

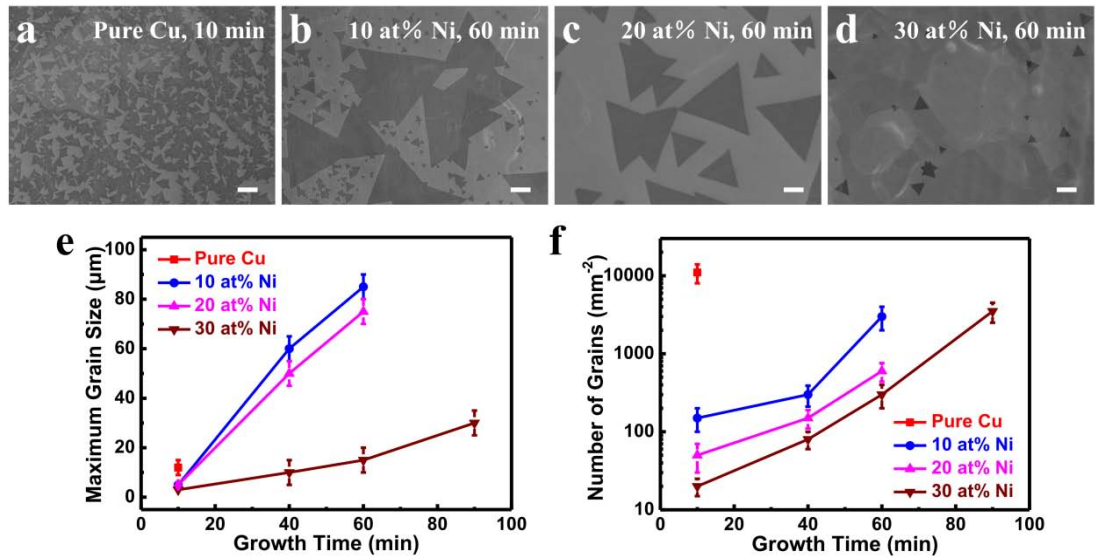

**Figure 2 | Effect of Ni concentration on the maximum grain size and the number of grains.** (**a**) SEM images of h-BN grains grown on Cu foil for 10 min at 1,050 °C. (**b-d**) SEM images of h-BN grains grown for 60 min on Cu-Ni alloy foils with (**b**) 10 atom %, (**c**) 20 atom % and (**d**) 30 atom % Ni at 1,050 °C, respectively. The scale bars are 20 μm in (**a-d**). (**e**) Dependence of the maximum h-BN grain size on Ni concentration and growth time. (**f**) Dependence of the number of h-BN grains on Ni concentration and growth time.

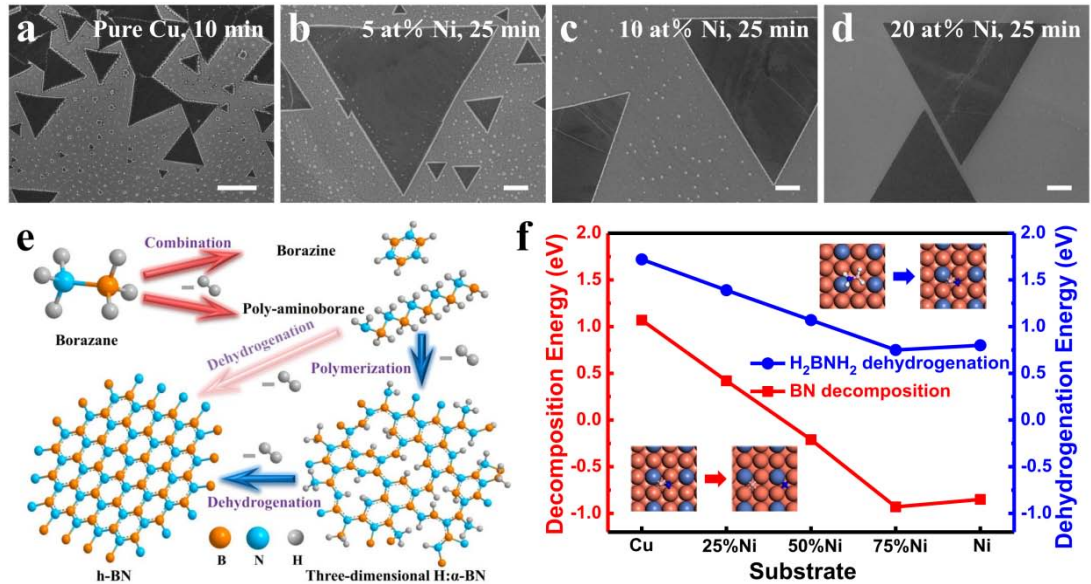

**Figure 3 | Enhancing the dissociation of borazane oligomers by introducing Ni.** (**a**) SEM images showing the morphology of a Cu foil after growth of h-BN for 10 min at 1,050 °C. (**b-d**) SEM images of Cu-Ni alloy foils with (**b**) 5 atom %, (**c**) 10 atom % and (**d**) 20 atom % Ni after growth of h-BN for 25 min at 1,050 °C, respectively. (**e**) Schematic diagram showing the structural development from dissociation of borazane to h-BN. (**f**) The molecular dynamic simulation of $H_2BNH_2$ dissociation on the surface of different substrates. The scale bars in (**a-d**) are 5 μm.

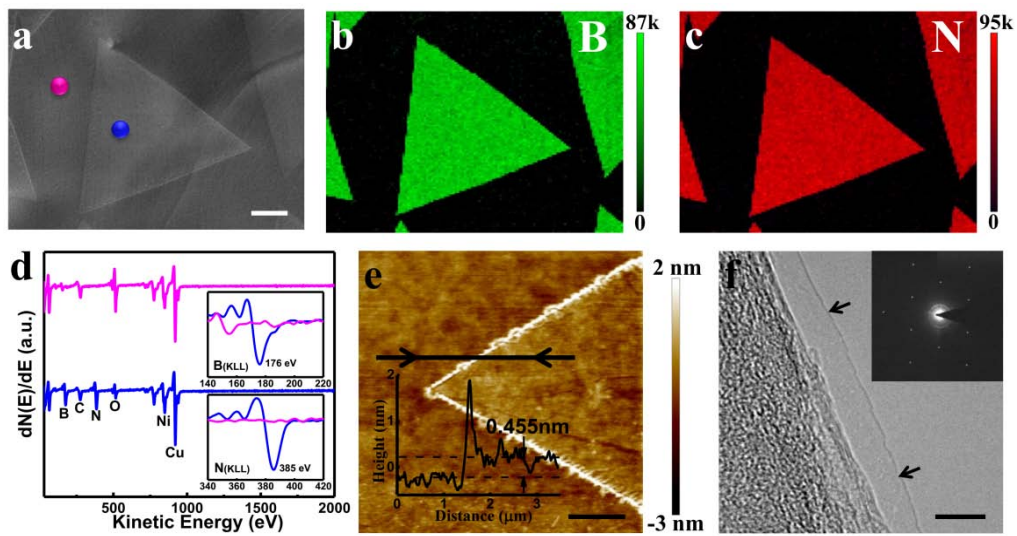

**Figure 4 | Characterization of h-BN grains.** (**a**) SEM image of a typical h-BN grain grown on a Cu-Ni alloy foil. (**b-c**) The corresponding B(KLL) and N(KLL) Auger electron maps, respectively. (**d**) AES spectra taken in the dotted areas shown in (**a**). The insets show the corresponding B(KLL) and N(KLL) spectra, respectively. (**e**) AFM image of a corner of a h-BN grain transferred on SiO$_2$/Si substrate. The inset shows the height distribution along the black line in (**e**). (**f**) HRTEM image of h-BN film on TEM grid with the black arrows showing that it is of monolayer thick. The inset shows the corresponding SAED pattern. Scale bars: (**a**) 10 μm, (**e**) 1 μm and (**f**) 5 nm.

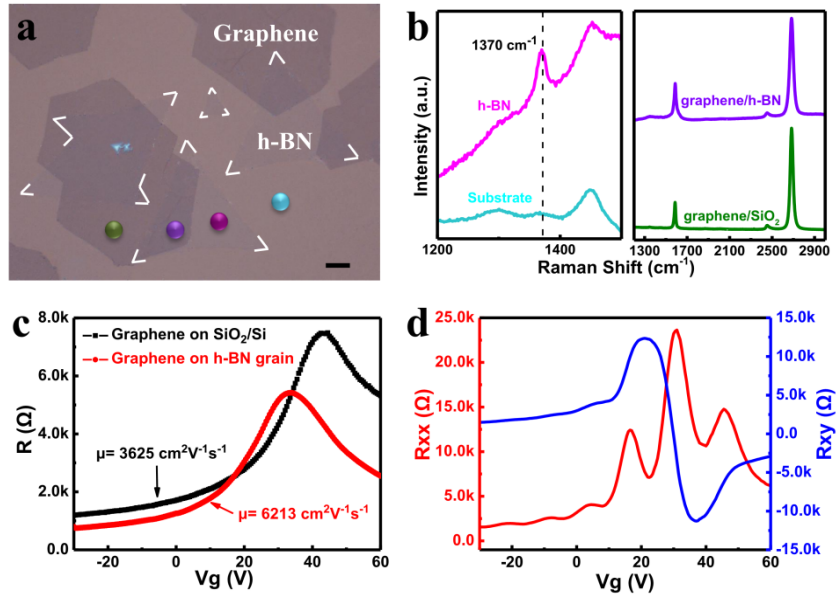

**Figure 5 | Characterization of graphene/h-BN hetero-structure.** (**a**) Optical image of graphene/h-BN grains on a 90 nm $SiO_2$/Si substrate. (**b**) Raman spectra taken from the dotted areas shown in (**a**). (**c**) Resistivity of graphene/$SiO_2$/Si and graphene/h-BN Hall bar devices versus the back gate voltage at room temperature, respectively. The values of the extracted carrier mobility are indicated in insets. (**d**) Graphene/h-BN Hall device exhibits half-integer quantum Hall effect measured at 4 K under a magnetic field $B = 9$ T. The scale bar in (**a**) is 10 μm.